\documentclass{nst}

\usepackage{subfigure,dcolumn}
\usepackage{epstopdf}
\usepackage{mhchem}
\usepackage{upgreek}
\usepackage{tabularx}

\begin{document}

\title{Producing $\mu$Bq/m$^3$-level low-$^{226}$Ra ultrapure water for the Jiangmen Underground Neutrino Observatory}
\thanks{This work was supported by the Strategic Priority Research Program of the Chinese Academy of Sciences (No. XDA10011200) and the Youth Innovation Promotion Association of the Chinese Academy of Sciences (No. 2023015).}

\author{Xi-Ting Xu}
\affiliation{Institute of Theoretical Physics, Shanxi University, Taiyuan, 030006, China}

\author{Cong Guo}
\email[Corresponding author, ]{C. Guo, E-mail address: guocong@ihep.ac.cn}
\affiliation{Experimental Physics Division, Institute of High Energy Physics, Chinese Academy of Sciences, Beijing, 100049, China}
\affiliation{School of Physics, University of Chinese Academy of Sciences, Beijing, 100049, China}
\affiliation{State Key Laboratory of Particle Detection and Electronics, Beijing, 100049, China}

\author{Jin-Chang Liu}
\affiliation{Experimental Physics Division, Institute of High Energy Physics, Chinese Academy of Sciences, Beijing, 100049, China}
\affiliation{School of Physics, University of Chinese Academy of Sciences, Beijing, 100049, China}
\affiliation{State Key Laboratory of Particle Detection and Electronics, Beijing, 100049, China}

\author{Yong-Peng Zhang}
\affiliation{Experimental Physics Division, Institute of High Energy Physics, Chinese Academy of Sciences, Beijing, 100049, China}
\affiliation{School of Physics, University of Chinese Academy of Sciences, Beijing, 100049, China}
\affiliation{State Key Laboratory of Particle Detection and Electronics, Beijing, 100049, China}

\author{Xiao-Hua Liang}
\affiliation{Astro-particle Physics Division, Institute of High Energy Physics, Chinese Academy of Sciences, Beijing, 100049, China}

\author{Xin-Jian Wen}
\email[Corresponding author, ]{X.J. Wen, E-mail address: wenxj@sxu.edu.cn}
\affiliation{Institute of Theoretical Physics, Shanxi University, Taiyuan, 030006, China}

\author{Chang-Gen Yang}
\affiliation{Experimental Physics Division, Institute of High Energy Physics, Chinese Academy of Sciences, Beijing, 100049, China}
\affiliation{School of Physics, University of Chinese Academy of Sciences, Beijing, 100049, China}
\affiliation{State Key Laboratory of Particle Detection and Electronics, Beijing, 100049, China}

\author{Quan Tang}
\email[Corresponding author, ]{Q. Tang, E-mail address: tangquan528@sina.com}
\affiliation{School of Nuclear Science and Technology, University of South China, Hengyang, 421001, China}

\author{Li-Dan Lv}
\affiliation{School of Nuclear Science and Technology, University of South China, Hengyang, 421001, China}

\author{Yuan-Hao Niu}
\affiliation{Institute of Theoretical Physics, Shanxi University, Taiyuan, 030006, China}

\author{Bin Xiao}
\affiliation{School of Mechanics and Optoelectronic Physics, Anhui University of Science and Technology, Huainan, 232001, China}

\begin{abstract}
The Jiangmen Underground Neutrino Observatory (JUNO), a 20~kton low-background liquid-scintillator (LS) detector, was designed primarily to determine the neutrino mass ordering. JUNO requires ultrapure water (UPW) with a $^{226}$Ra concentration $<$50~$\mu$Bq/m$^3$ because of the direct liquid-liquid contact between LS and UPW during detector filling. To meet this stringent requirement, a high-sensitivity measurement system capable of detecting $^{226}$Ra in water down to 3.9~$\mu$Bq/m$^3$ was developed, and the 100~t/h UPW production process was optimized. By integrating selective ion-exchange resin with membrane-based separation technologies, the system consistently produced UPW with a $^{226}$Ra concentration of $<$4~$\mu$Bq/m$^3$, exceeding the JUNO specification and, to our knowledge, representing one of the lowest reported levels achieved at this engineering scale. This paper describes the design and implementation of JUNO's UPW system and the high-sensitivity $^{226}$Ra measurement platform, together with a systematic evaluation of $^{226}$Ra removal efficiencies across the purification train and validation of the final water quality.
\end{abstract}


\maketitle

\section{Introduction}\label{sec.I}

The Jiangmen Underground Neutrino Observatory (JUNO)~\cite{bib1,bib2,bib2-0-1,bib2-0-2} is a multipurpose experiment designed primarily to determine the neutrino mass ordering and to measure with high precision neutrino oscillation parameters by detecting reactor antineutrinos. With 20~ktons of low-background liquid-scintillator (LS) and an unprecedented energy resolution of 3\%@1MeV, JUNO is the largest underground LS detector and can address a broad range of key questions in astroparticle physics. JUNO's extensive physics program encompasses supernova neutrinos, atmospheric neutrinos, solar neutrinos, geoneutrinos, and searches for new physics~\cite{bib1,bib2}. Located in an underground laboratory with 700~m overburden, the muon flux at the JUNO site is about 4~mHz/m$^2$ with a mean energy of 207~GeV~\cite{bib2,bib3}. The primary detection channel for reactor antineutrinos in JUNO is the inverse beta decay reaction on protons. JUNO is expected to detect only $\sim$60 reactor antineutrino events per day, making stringent control of radioactive backgrounds essential~\cite{bib4}.

\begin{figure*}[htb]
\includegraphics[width=13cm]{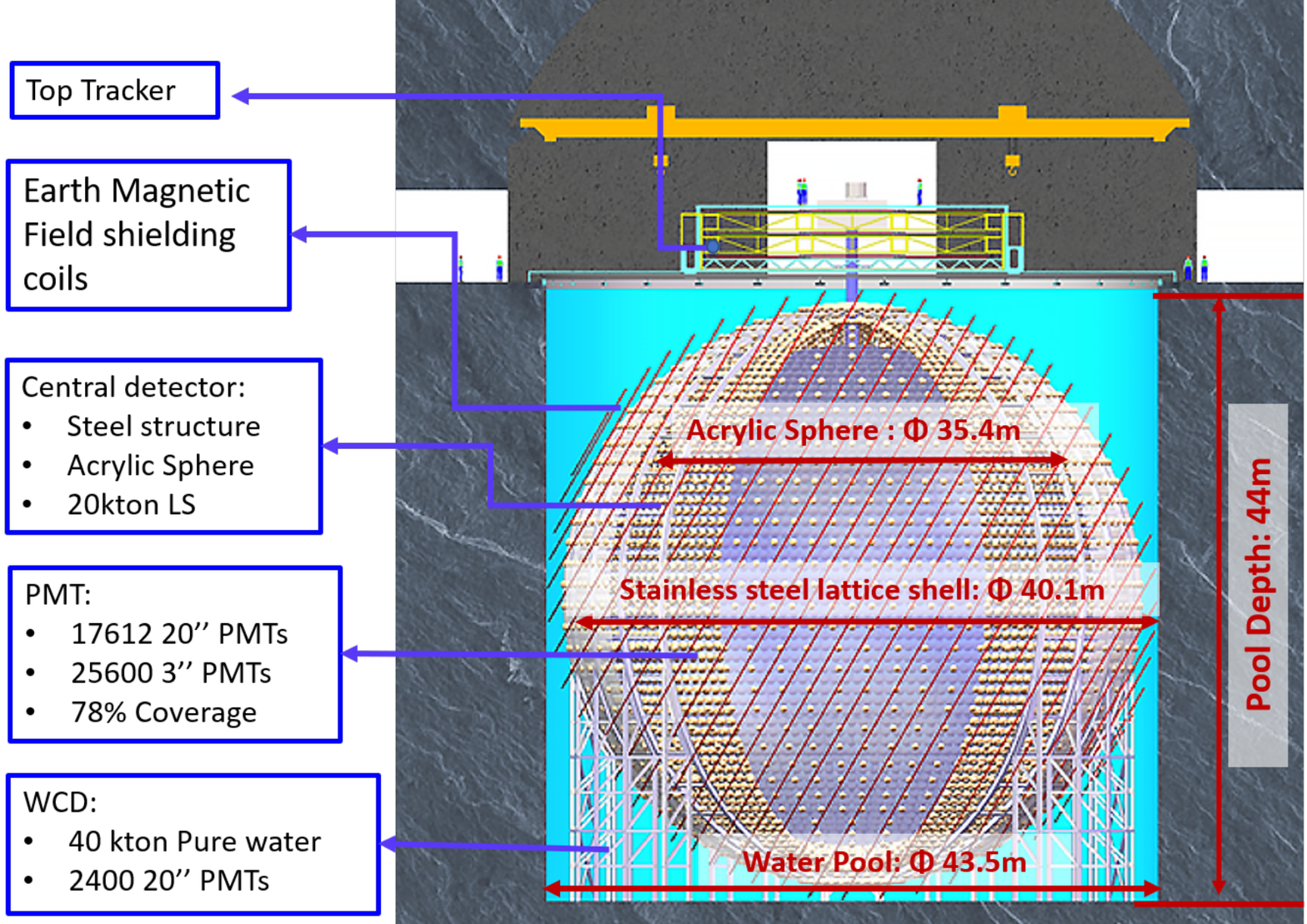}
\caption{Schematic drawing of the JUNO detector showing the main components of the experiment.}
\label{fig.detector}
\end{figure*}

Fig.~\ref{fig.detector} presents a schematic view of the JUNO detector, which consists of a Central Detector (CD), a Water Cherenkov Detector (WCD), and a Top Tracker (TT) detector. The CD contains 20~kton of ultrapure liquid scintillator (LS) housed in a spherical acrylic vessel and is submerged in the WCD, which holds 40~ktons of ultrapure water (UPW) and is instrumented with 2400 20-inch microchannel-plate photomultiplier tubes (MCP-PMTs). The WCD provides sufficient water shielding to protect the CD from ambient radioactivity and simultaneously serves as a cosmic muon veto. For background and safety reasons, the CD and WCD must be initially filled simultaneously with UPW. Subsequently, the UPW in the CD is replaced with LS at a production rate of 7~m$^3$/h. During this replacement process, UPW and LS come into direct contact. As this stage marks the final step before detector operation, strict control of radionuclides in water is critical to prevent contamination of the LS by the UPW. One of the most challenging requirements is to maintain the $^{226}$Ra concentration in the water to below 50~$\mu$Bq/m$^3$~\cite{bib2}.

In the field of low-background water purification and $^{226}$Ra detection, the Sudbury Neutrino Observatory (SNO)~\cite{bib2-2} and Super-Kamiokande (Super-K)~\cite{bib2-3} experiments have achieved internationally recognized advances. SNO developed a measurement technique using manganese oxide-coated beads, achieving a $^{226}$Ra detection sensitivity of $\sim$10~$\mu$Bq/m$^3$. The initial $^{226}$Ra concentration in SNO's UPW was $\sim$50~$\mu$Bq/m$^3$, which was later reduced to 10--20~$\mu$Bq/m$^3$ through recirculation and purification~\cite{bib14}. Super-K adopted SNO's methodology and attained a similar sensitivity of $\sim$10~$\mu$Bq/m$^3$, although its UPW exhibited slightly higher $^{226}$Ra levels, with an initial concentration of 61 $\pm$ 13~$\mu$Bq/m$^3$, decreasing to 32 $\pm$ 7~$\mu$Bq/m$^3$ after recirculation~\cite{bib2-4}. In terms of production scale, Super-K's water system has a capacity of 30~t/h~\cite{bib2-4}, substantially larger than SNO's 9~t/h system~\cite{bib2-2}.

In this work, we report a large-scale UPW purification system that reduces the $^{226}$Ra concentration from $\sim$4 $\times$ 10$^5$~$\mu$Bq/m$^3$ in the source water to $<$4~$\mu$Bq/m$^3$, surpassing both the JUNO requirement and the performance of comparable experiments. To achieve such ultralow detection, we developed a high-sensitivity measurement system based on Mn-fiber adsorption and radon emanation, reaching a sensitivity of 3.9~$\mu$Bq/m$^3$ through optimized Mn-fiber synthesis. The system operates reliably at 100~t/h, demonstrating its feasibility for large-scale engineering. Furthermore, through systematic optimization and pollution source tracing, we identified the softening resin and reverse osmosis membrane as the key removal components (both achieving $>$99\% radium removal efficiency), and traced the dominant recontamination sources to sediment introduced by the regenerant salt and to air ingress through tank breather valves, thereby enabling targeted engineering improvements for long-term stability.

This paper is structured as follows. Sec.~\ref{sec.UPW} describes the UPW production system used for the detector water inventory. Sec.~\ref{sec.Rameasurement} details the high-sensitivity $^{226}$Ra-in-water measurement system with a 3.9~$\mu$Bq/m$^3$ detection sensitivity. The measurement results of the $^{226}$Ra concentrations after each purification stage and in the final product water are presented in Sec.~\ref{sec.Results}. Potential upgrades to the system for enhanced $^{226}$Ra removal are discussed in Sec.~\ref{sec.discussion}, and a summary of the work is provided in Sec.~\ref{sec.summary}.

\section{UPW system}\label{sec.UPW}

The JUNO experiment uses a 100~t/h UPW system to supply 40~ktons of UPW for the WCD. This system produces water that meets the following specifications: a resistivity of $>$18.0~M$\Omega\cdot$cm, a $^{222}$Rn concentration of $<$1 $\times$ 10$^4$~$\mu$Bq/m$^3$, and a controlled temperature of 17~$^{\circ}$C, thereby ensuring high muon detection efficiency for the WCD, negligible radioactive contribution to the CD, and thermal stability for the CD~\cite{bib1}. In addition, for the initial filling of the acrylic vessel, this system must deliver UPW at 50~t/h with a $^{226}$Ra concentration below 50~$\mu$Bq/m$^3$~\cite{bib2}.

\begin{figure*}[htb]
\includegraphics[width=16cm]{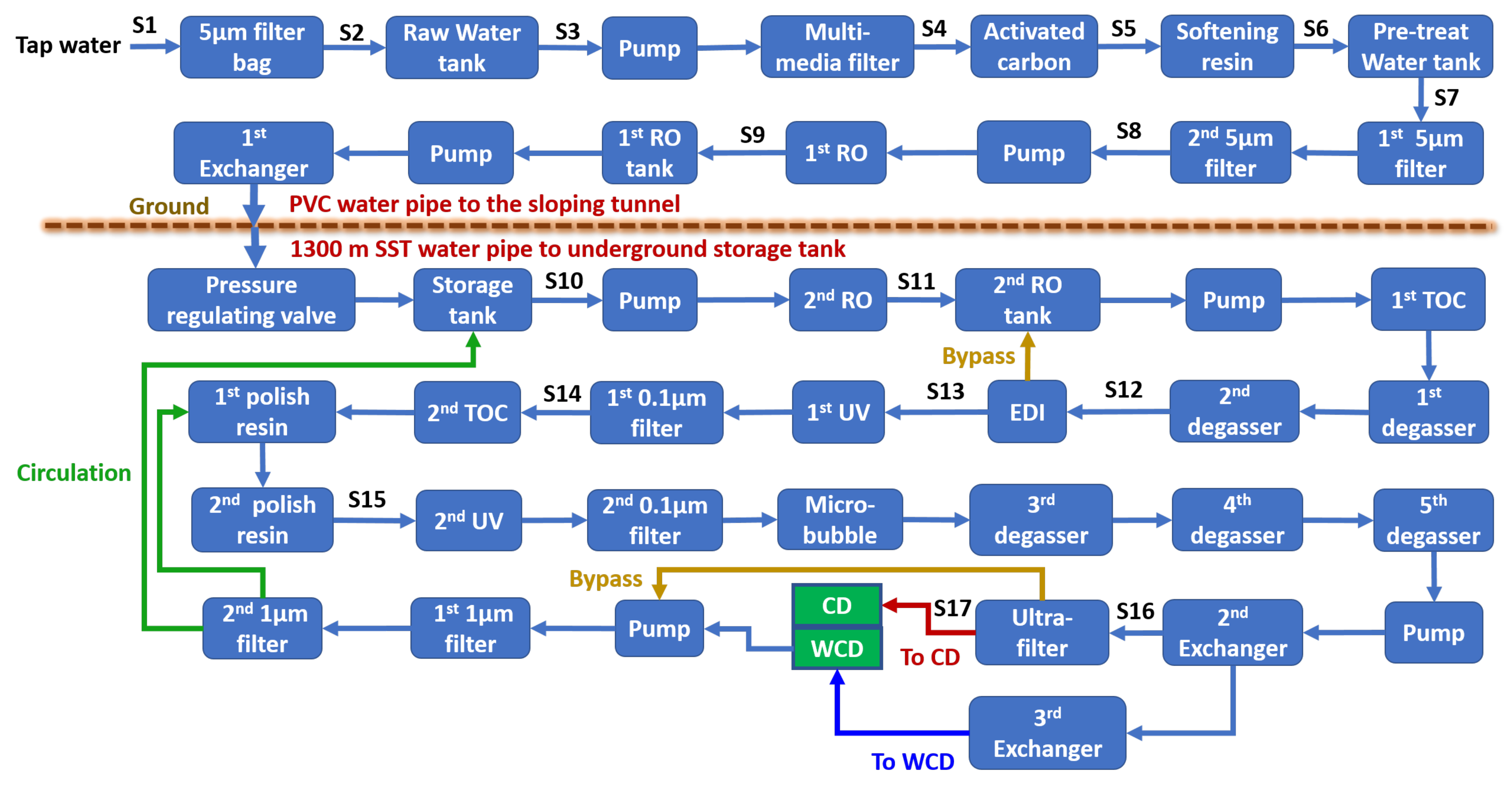}
\caption{The flow chart of the JUNO 100~t/h UPW system.}
\label{fig.watersystem}
\end{figure*}

The JUNO 100~t/h UPW system uses tap water as feed-water to produce UPW that is compliant with the Chinese national EW-I electronic-grade water standard~\cite{bib5}. As shown in Fig.~\ref{fig.watersystem}, the system comprises aboveground and underground sections. The aboveground treatment system consists of the following sequential units: \\
\indent- Bag filters (pore size: 5~$\mu$m).\\
\indent- Multi-media filters (with layered quartz sand, activated carbon, and anthracite). \\
\indent- Activated carbon filters (using coconut-shell-derived carbon).\\
\indent- Softening resin columns (with a dedicated $^{226}$Ra removal cation-exchange resin from Jiangsu Suqing Water Treatment Engineering Group Co., Ltd.).\\
\indent- Cartridge filters (pore size: 5~$\mu$m).\\
\indent- Reverse Osmosis (RO) membranes (Model TM720D-400, Toray Industries.).

The pretreatment process begins with mechanical filtration using bag filters for bulk particulate removal, followed by multi-media filtration to reduce suspended solids. Subsequent activated carbon adsorption removes residual chlorine, organic compounds, ammoniacal nitrogen, nitrites, trace heavy metals, and microbial contaminants. The ion-exchange softening process removes Ca$^{2+}$ and Mg$^{2+}$ through cation exchange. After intermediate storage, water passes through cartridge filters (5~$\mu$m) to protect downstream equipment. The RO array provides primary desalination via semipermeable membranes, achieving near-complete ionic rejection~\cite{bib6}.

The multi-media filters, activated carbon units, and softening resins are regenerable treatment units. Multi-media filters and activated carbon filters undergo hydraulic backwashing to restore capacity, while softening resins are regenerated with NaCl brine to displace accumulated ions and restore exchange capacity. This approach ensures sustainable long-term operation.

The product water from the aboveground system is transferred underground by a 1300-m long stainless-steel pipeline. With an elevation differential of $\sim$450~m, the water first passes through a pressure-reducing valve before entering the storage tank. The underground water system comprises the following components:\\
\indent- Secondary RO membranes (for enhanced desalination).\\
\indent- Electrodeionization (EDI) unit.\\
\indent- Cartridge filters (with a pore size of 0.1~$\mu$m).\\
\indent- Total Organic Carbon (TOC) removal unit.\\
\indent- Ultraviolet (UV) sterilizers.\\
\indent- Degassing membranes.\\
\indent- Microbubble generator.\\
\indent- Heat exchangers.\\
\indent- Ultrafiltration (UF, Model NTU-3306-K6R-HP, Nitto Denko Corporation).

The secondary RO membranes provide additional ion removal. The EDI unit combines ion-selective membranes and resin under a DC field for continuous deionization without chemical regeneration~\cite{bib7}. Cartridge filters capture residual particles. The TOC/UV units sterilize the water by UV irradiation while simultaneously oxidizing residual organic compounds. Degassing membranes, which employ hollow fibers, vacuum pumps, and purging nitrogen, remove dissolved gases via partial-pressure differentials and are critical for stripping $^{222}$Rn from water~\cite{bib8,bib9,bib10}. The microbubble generator enhances the $^{222}$Rn removal efficiency of the degassing membranes by reintroducing nitrogen into the water~\cite{bib8,bib10}. Heat exchangers regulate water temperature. UF membranes (with a pore size of $<$10~nm) remove fine particles, colloids, macromolecular impurities, and bacteria from the water.

The 100~t/h UPW system supplies water for both the WCD and the initial filling of the CD. Owing to their distinct requirements, the two streams are treated separately. The CD-bound water, which contacts the LS directly, receives additional UF for enhanced removal of particulates and colloid-associated radionuclides ($^{238}$U/$^{232}$Th/$^{226}$Ra). The WCD-bound water, by contrast, is routed through a dedicated heat exchanger to remove heat from the submerged PMT electronics, thereby ensuring stable detector temperatures.

\section{$^{226}$Ra concentration in water measurement}\label{sec.Rameasurement}

The required $^{226}$Ra concentration of 50~$\mu$Bq/m$^3$ ($\sim$1.4 $\times$ 10$^{-21}$~g/g) is far below the direct detection capability of conventional techniques such as high-purity germanium $\gamma$ spectrometry~\cite{bib11} and inductively coupled plasma mass spectrometry~\cite{bib12}. Therefore, a custom system was developed that determines the $^{226}$Ra activity by measuring its gaseous daughter, $^{222}$Rn.

\subsection{Measurement principle}
The first step in measuring the $^{226}$Ra concentration in water via $^{222}$Rn emanation is the preconcentration of $^{226}$Ra. This method exploits the strong adsorption of $^{226}$Ra onto MnO$_2$~\cite{bib13,bib14,bib15} by using laboratory-synthesized manganese fiber (Mn-fiber)~\cite{bib16,bib17}.

\begin{figure}[htb]
\includegraphics[width=7.5cm]{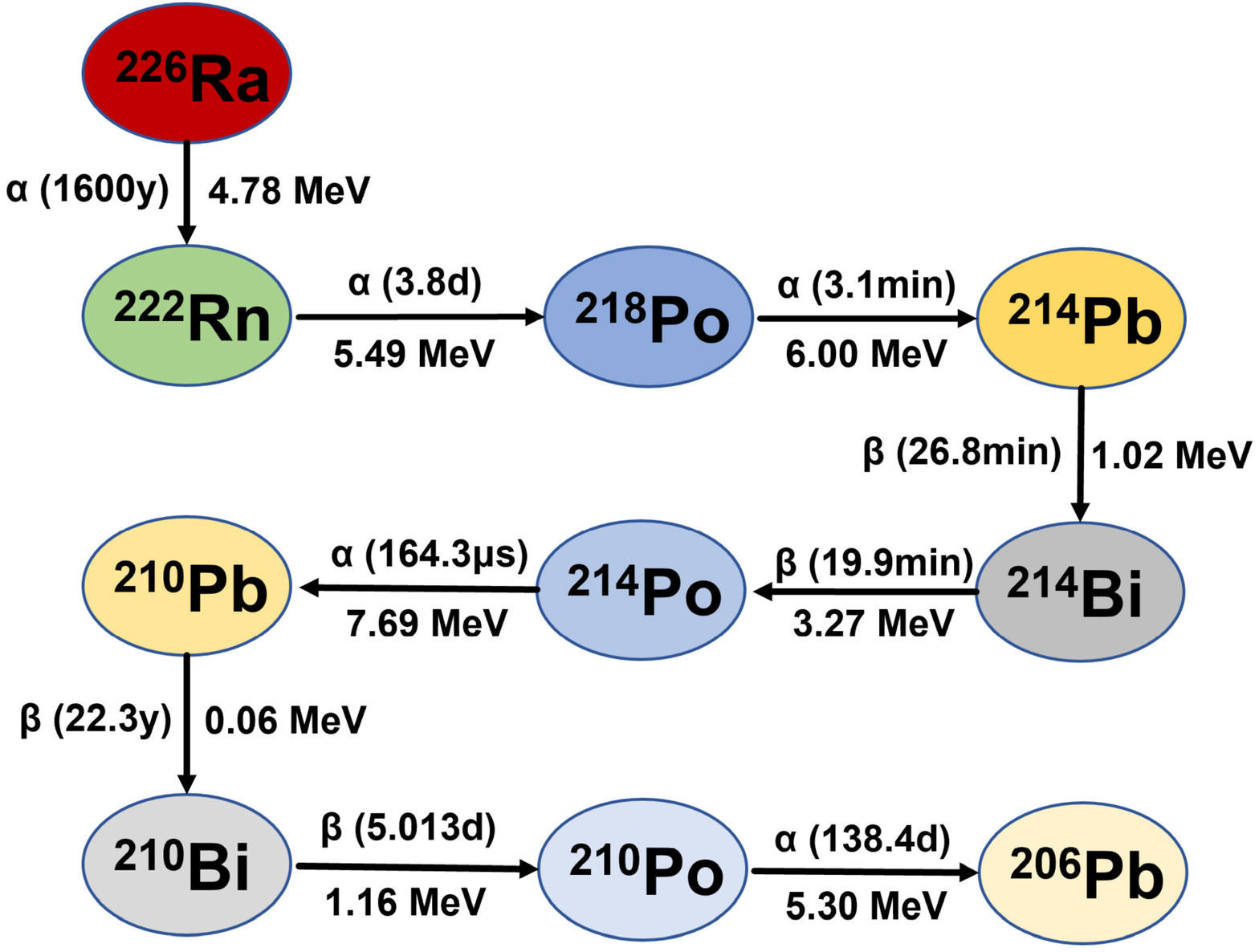}
\caption{The relevant decay branches of $^{226}$Ra. The times listed in parentheses are the half-lives. The energies shown for $\alpha$ decay are peak energies, while those for $\beta$ decay are the maximum energies.}
\label{fig.decaychain}
\end{figure}

Following $^{226}$Ra extraction, the activity of $^{222}$Rn emanated from the Mn-fiber during a sealed period is measured. Fig.~\ref{fig.decaychain} illustrates the relevant decay branch of $^{226}$Ra. During sealing, the relationship between the activity of $^{226}$Ra and its $^{222}$Rn progeny activity can be calculated using Eq.~\ref{eq1}~\cite{bib16}:
\begin{equation}
\label{eq1}
A_{\mathrm{Rn}}(t) = A_{\mathrm{Ra}} \times \frac{\lambda_{\mathrm{Rn}}}{\lambda_{\mathrm{Rn}} - \lambda_{\mathrm{Ra}}} \times  ( e^{-\lambda_{\mathrm{Ra}} t} - e^{-\lambda_{\mathrm{Rn}} t} )
\end{equation}
where $A_{\mathrm{Rn}}(t)$ is the $^{222}$Rn activity after a sealing time of $t$ in mBq, $A_{\mathrm{Ra}}$ is the initial $^{226}$Ra activity adsorbed on the Mn-fiber in mBq, $\lambda_{\mathrm{Rn}}$ is the decay constant of $^{222}$Rn, $\lambda_{\mathrm{Ra}}$ is the decay constant of $^{226}$Ra, and $t$ is the sealing time in hours.

Given that the half-life of $^{226}$Ra is significantly longer than that of $^{222}$Rn (i.e., $\lambda_{\mathrm{Ra}} \ll \lambda_{\mathrm{Rn}}$), Eq.~\ref{eq1} can be simplified to Eq.~\ref{eq2}:
\begin{equation}
\label{eq2}
A_{\mathrm{Ra}} = \frac{A_{\mathrm{Rn}}(t)}{1-e^{-\lambda_{\mathrm{Rn}} t}}
\end{equation}

\subsection{Measurement procedures and devices}
The $^{226}$Ra concentration measurement consists of four sequential steps: (1) Mn-fiber synthesis, (2) $^{226}$Ra extraction from water, (3) Mn-fiber encapsulation, and (4) $^{222}$Rn activity determination.

\subsubsection{Mn-fiber}

To meet JUNO's requirement for $\mu$Bq/m$^3$-level $^{226}$Ra measurements, we significantly reduced the $^{226}$Ra background of the key adsorbent, the Mn-fiber, from $\sim$17~$\mu$Bq/g to 10.7 $\pm$ 1.5~$\mu$Bq/g. This was achieved through three major modifications to the standard synthesis protocol~\cite{bib16,bib17}:

(1) Nitric-acid precleaning of the polyurethane fibers.

(2) Relocating the boiling and rinsing steps to a cleanroom.

(3) Replacing distilled water with commercially supplied high-purity water from China Resources C'estbon Beverage (China) Co., Ltd. during rinsing---the most effective modification.

\begin{equation}
\label{eq3}
\ce{2KMnO4 + 2H2SO4 ->[\Delta] Mn2O7 + 2KHSO4 + H2O}
\end{equation}
\begin{equation}
\label{eq4}
\ce{2Mn2O7 -> 4MnO2 + 3O2 ^}
\end{equation}

\begin{figure}[htb]
\includegraphics[width=8cm]{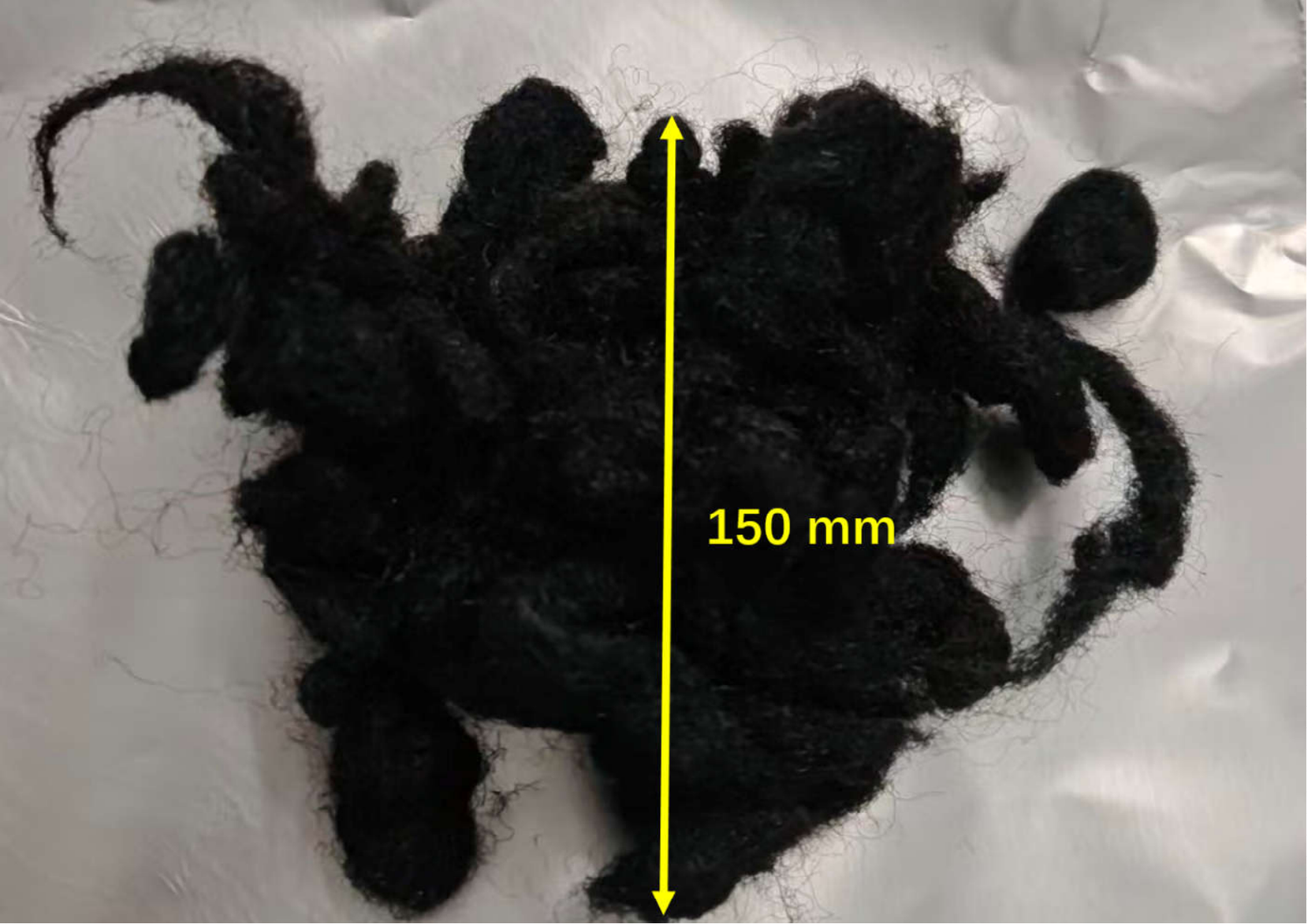}
\caption{A picture of the laboratory-synthesized Mn-fiber.}
\label{fig.Mnfiber}
\end{figure}
The Mn-fiber itself is prepared by forming an adherent MnO$_2$ coating on polyurethane fibers through the reaction of KMnO$_4$ with H$_2$SO$_4$ under heating (Eq.~\ref{eq3}, Eq.~\ref{eq4}). An image of the Mn-fiber is shown in Fig.~\ref{fig.Mnfiber}. Gravimetric analysis indicates a $\sim$13\% mass gain of the polyurethane fibers after conversion to Mn-fiber.

\subsubsection{Extraction column}

\begin{figure}[htb]
\includegraphics[width=6cm,height=8cm]{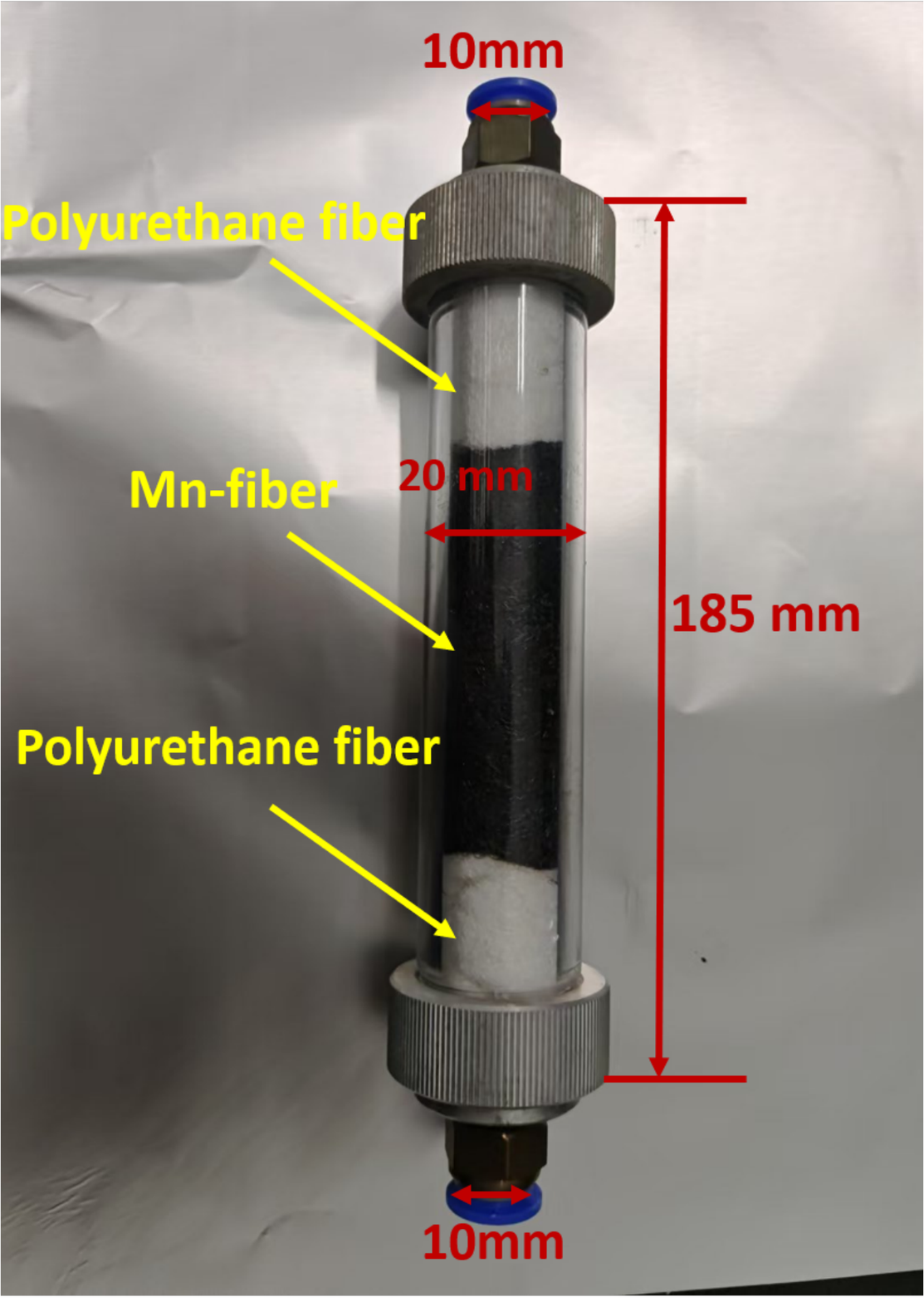}
\caption{A picture of the column packed with Mn-fiber.}
\label{fig.column}
\end{figure}

The $^{226}$Ra extraction column, illustrated in Fig.~\ref{fig.column}, is fabricated from an acrylic tube ($\sim$20~mm inner diameter and $\sim$185~mm long) sealed with rubber gaskets and 10~mm quick-connects. The assembled unit tolerates water pressures up to $\sim$0.5~MPa. Approximately 5~g of Mn-fiber is packed in the central section, flanked by polyurethane fibers at both ends to retain particulate matter. During operation, water is directed upward through the column to maximize contact with the Mn-fiber. A residential water meter ($\pm$2\% accuracy, Model 099Q, AMICO) at the outlet records the total volume of water passed through the Mn-fiber.

\subsubsection{$^{222}$Rn emanation measurement system}

\begin{figure*}[htb]
\includegraphics[width=13cm]{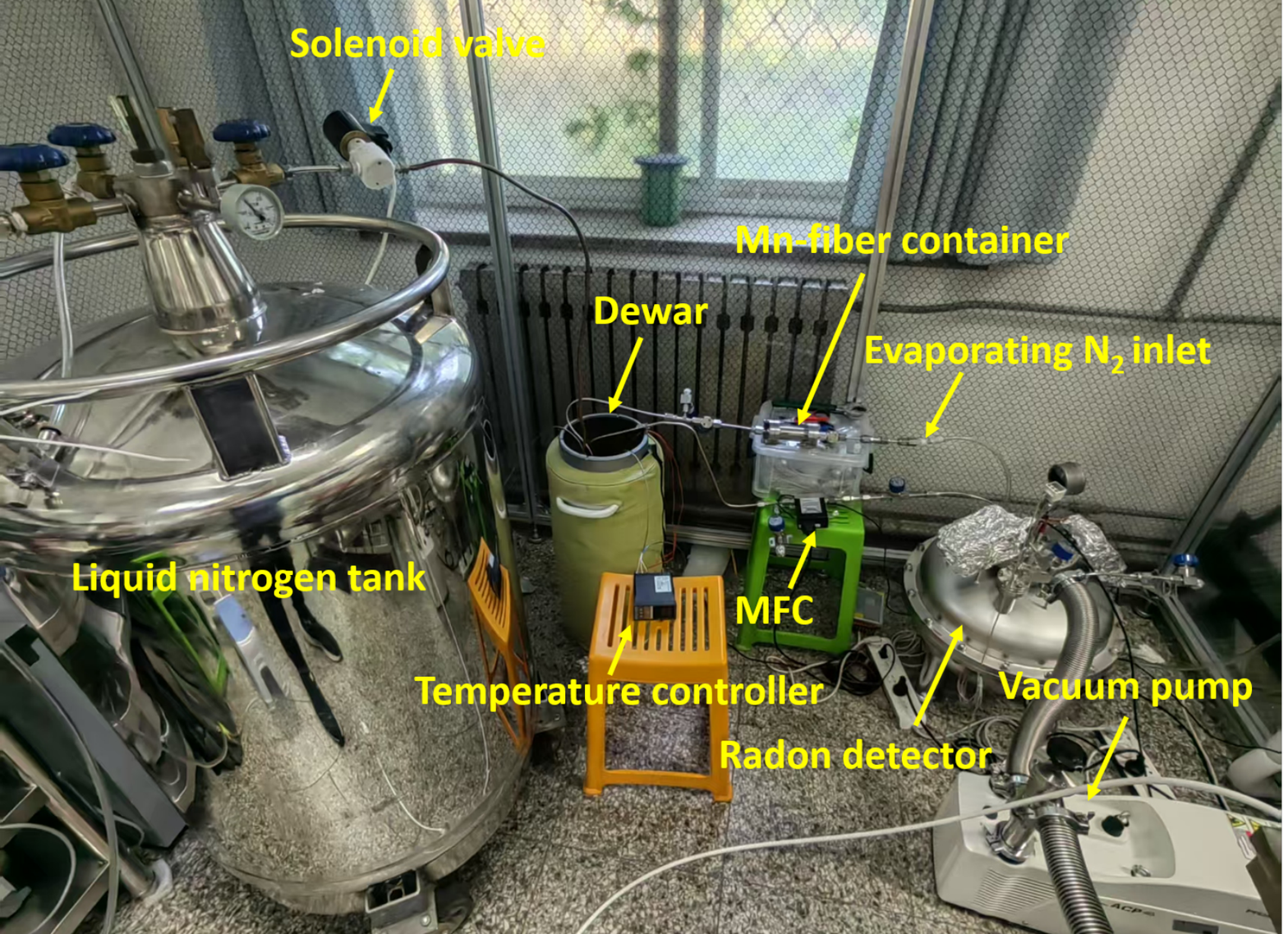}
\caption{A picture of the radon emanation measurement system.}
\label{fig.measurementsystem}
\end{figure*}

The $^{226}$Ra activity extracted by Mn-fibers was quantified by measuring its gaseous daughter, $^{222}$Rn, using a radon-emanation measurement system (Fig.~\ref{fig.measurementsystem}). The system comprises a radon detector, a Mn-fiber container, a dehumidification system, a mass flow controller (MFC, Model 1179A, MKS), and a vacuum pump (Model ACP40, Pfeiffer Vacuum). All connections use metal-sealed Con-Flat flanges and VCR components, ensuring a leak rate below $1 \times 10^{-9}$~Pa$\cdot$m$^3$/s. Key components are summarized below; full details are available in our previous work~\cite{bib17}.

(1) Radon detector. The radon detector operates on the principle of electrostatic collection~\cite{bib9,bib19,bib20,bib21,bib22,bib23,bib24,bib25}, where the detection efficiency for radon progeny depends on the applied collection voltage and gas humidity. At 3\% relative humidity and $\sim$700~V collection voltage, the detector achieves 90\% collection efficiency for $^{222}$Rn progeny $^{214}$Po. This corresponds to a calibration factor ($C_F$) of 67 $\pm$ 6.7 counts per hour per Bq/m$^3$ (CPH/(Bq/m$^3$)). The system's single-day measurement sensitivity is 0.7~mBq/m$^3$ for $^{222}$Rn~\cite{bib19}. For each sample measurement, the radon detector accumulates counts over a 24-h data-acquisition period. As the measured value represents the average $^{222}$Rn concentration during this interval, determination of the initial $^{222}$Rn concentration requires correction for radioactive decay, which can be calculated using Eq.~\ref{eq5}~\cite{bib17}:
\begin{equation}
\label{eq5}
A_{\mathrm{Rn}} = \frac{A_{\mathrm{Rn-M}} \times \lambda_{\mathrm{Rn}} \times (t_2 - t_1)}{e^{-\lambda_{\mathrm{Rn}} t_1}-e^{-\lambda_{\mathrm{Rn}} t_2}}
\end{equation}
where $A_{\mathrm{Rn}}$ is the initial $^{222}$Rn activity after transfer into the detector in mBq, $A_{\mathrm{Rn-M}}$ is the measured $^{222}$Rn activity in mBq, $t_1$ is the measurement start time in hours, and $t_2$ is the measurement end time in hours of the counting interval.

\begin{figure}[htb]
\includegraphics[width=7.5cm]{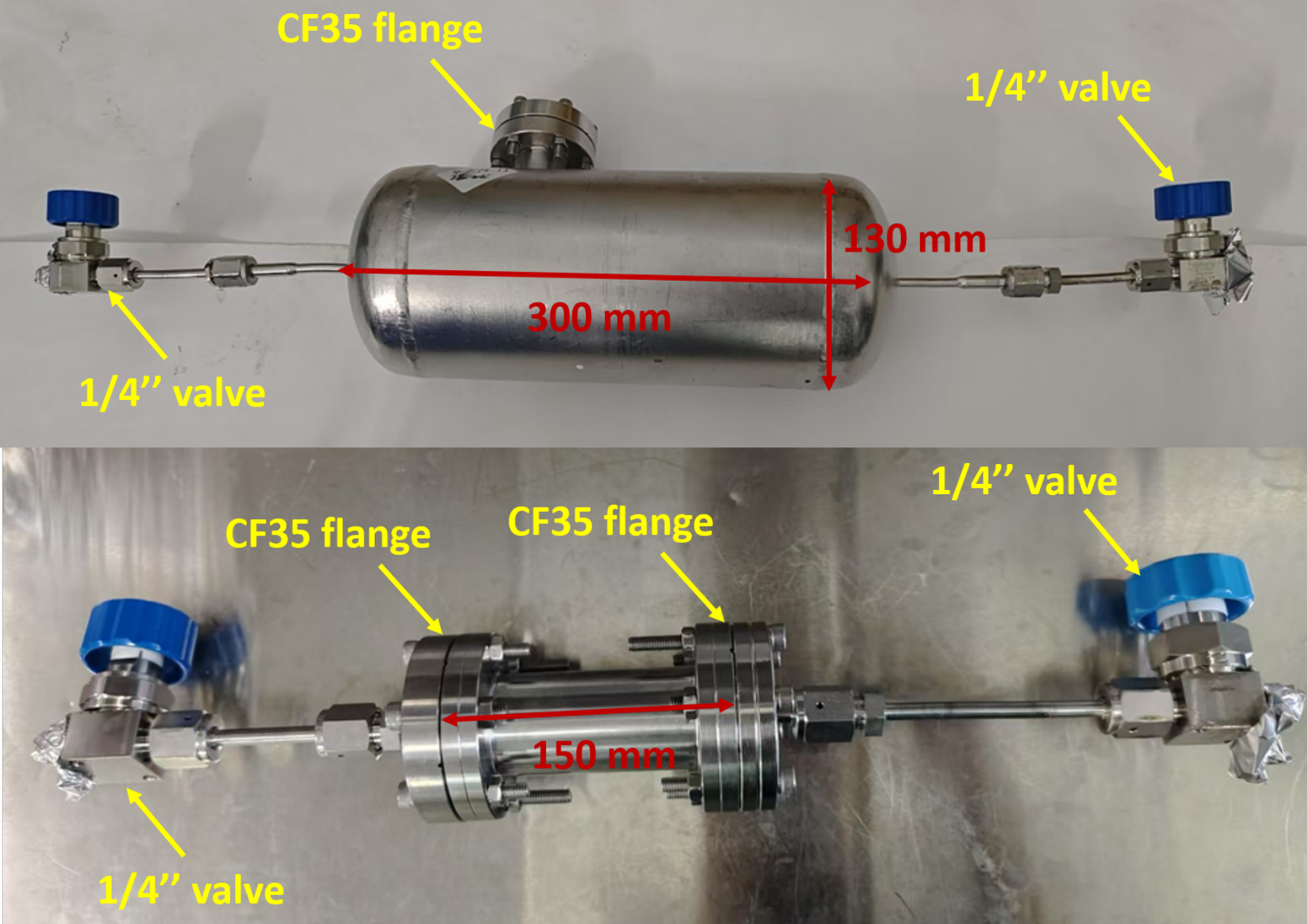}
\caption{Pictures of the Mn-fiber containers used in this work.}
\label{fig.container}
\end{figure}

(2) Mn-fiber container. Two chamber types are used (Fig.~\ref{fig.container}): a standard CF35 tube ($\sim$150~mm long) for 5~g samples, and a custom chamber ($\sim$130~mm diameter and $\sim$300~mm length) for background measurements with sample masses below 300~g. All inner surfaces were electropolished to a roughness of $<$0.1~$\mu$m. After loading, chambers are purged with boil-off nitrogen to remove residual air and then sealed for several days to allow $^{222}$Rn ingrowth. The activity ratio between $^{226}$Ra and $^{222}$Rn can be calculated using Eq.~\ref{eq2}. For instance, after a 10-day sealing interval, $^{222}$Rn activity reaches 83.7\% of the initial $^{226}$Ra activity.

(3) Dehumidification system. Mn-fibers that have extracted $^{226}$Ra retain substantial water, releasing significant water vapor during radon transfer that reduces detector efficiency because of humidity effects. This system comprises a liquid nitrogen tank, a solenoid valve, a Dewar, a PT100 temperature sensor, and a temperature controller. Through the thermostatic regulation of intermittent liquid nitrogen injection into the Dewar, the system maintains a temperature of $-60^\circ$C inside the dewar, reducing the relative humidity to 3\% at a 1~L/min gas flow rate without introducing background.

(4) MFC. The MFC is used to regulate the gas flow rate during radon transfer operations, with an adjustment range of 0.5--5~L/min.

(5) Vacuum pump. Before measurements, the pump evacuates the radon detector and associated pipelines. Subsequently, boil-off nitrogen is used to purge the emanated $^{222}$Rn from the Mn-fiber container into the radon detector.

(6) Pipelines and valves. All stainless-steel gas pipelines and interconnecting valves are electropolished components.

\subsection{Extraction efficiency calibration}
The efficiency of extracting $^{226}$Ra from water with Mn-fiber was calibrated using a standardized $^{226}$Ra solution with a concentration of (9.04 $\pm$ 0.53) $\times$ 10$^3$~$\mu$Bq/m$^3$. Our prior work~\cite{bib17} showed that the extraction efficiency remains stable over $^{226}$Ra concentrations of 760--6.1 $\times$ 10$^4$~$\mu$Bq/m$^3$ and water flow rates of 1.6--8~L/min. To cover $^{226}$Ra concentrations from 1 to 1 $\times$ 10$^4$~$\mu$Bq/m$^3$ in JUNO, sample volumes are varied, which requires dedicated calibration of the extraction efficiency as a function of treated water volume. In contrast to absolute quantification, we employed a relative scheme with volume-matched control experiments to simultaneously reduce detector-systematic uncertainties and background contributions.

Tab.~\ref{tab.volume} presents the calibrated $^{226}$Ra extraction efficiencies across different water volumes. Throughout the test, an additional 150~$\mu$L of $^{226}$Ra solution was spiked into each 300~L water batch, and the water flow rate through the Mn-fiber was maintained at $\sim$5~L/min.

\begin{table}[!htb]
\caption{$^{226}$Ra extraction efficiencies of Mn-fiber for different water volumes. The uncertainties in UPW volume are derived from the accuracy of the water flow meter. The statistical uncertainties in efficiency are based on Poisson statistics, while the systematic uncertainties are determined by the uncertainty in the radium solution's activity. For samples No.~1--No.~4, the radon detector data-taking time was 24~h, whereas for samples No.~5 and No.~6, it was 15~h.}
\label{tab.volume}
\renewcommand{\arraystretch}{1.3}
\begin{tabular*} {7.4cm} {c|c|c}
\toprule
{No.}&{UPW volume (m$^3$) }&{ $^{226}$Ra extraction efficiency (\%)} \\ \hline
1  & 0.3 $\pm$ 0.006 & 99.0 $\pm$ 13.8$_{stat}$ $\pm$ 5.8$_{sys}$\\ \hline
2  & 0.5 $\pm$ 0.01  & 98.5 $\pm$ 10.6$_{stat}$ $\pm$ 5.8$_{sys}$ \\ \hline
3  & 1.0 $\pm$ 0.02  & 99.5 $\pm$ 7.6$_{stat}$ $\pm$ 5.8$_{sys}$\\ \hline
4  & 5.0 $\pm$ 0.1 & 94.6 $\pm$ 3.3$_{stat}$ $\pm$ 5.5$_{sys}$\\ \hline
5  & 10.0  $\pm$ 0.2  & 91.3 $\pm$ 2.9$_{stat}$ $\pm$ 5.4$_{sys}$\\ \hline
6  & 20.0 $\pm$ 0.4  & 89.2 $\pm$ 2.0$_{stat}$ $\pm$ 5.2$_{sys}$ \\
\bottomrule
\end{tabular*}
\end{table}

For water volumes below 1~m$^3$ passing through the Mn-fiber extraction column, $^{226}$Ra extraction efficiency remained close to $\sim$100\%. Beyond this volume, a progressive decline in adsorption efficiency was observed, likely reflecting partial desorption of $^{226}$Ra or shedding of the MnO$_2$ coating.

\subsection{Sensitivity estimation}
Based on the normal approximation of the background counts, the sensitivity of measuring $^{226}$Ra concentration in water at a 90\% confidence level (two-sided) can be estimated from Eq.~\ref{eq6}~\cite{bib24,bib25-2}:
\begin{equation}
\label{eq6}
L = \frac{1.645 \times \sigma_{BG} \times V_s}{24 \times C_F \times V_w \times \varepsilon \times 10^{-6}}
\end{equation}
where $L$ is the sensitivity in $\mu$Bq/m$^3$, $\sigma_{BG}$ is the statistical uncertainty of the system's background event rate in Counts Per Day (CPD), $V_s$ = 0.042~m$^3$ is the total volume of the measurement system, $C_F$ = 67 $\pm$ 6.7~CPH/(Bq/m$^3$) is the detector calibration factor, $V_w$ is the sample water volume in m$^3$, and $\varepsilon$ is the $^{226}$Ra extraction efficiency listed in Tab.~\ref{tab.volume}. This calculation uses the measured background variance together with the radon-ingrowth relation of Eq.~\ref{eq2} for the selected sealing time.

\begin{figure}[htb]
\includegraphics[width=7.5cm]{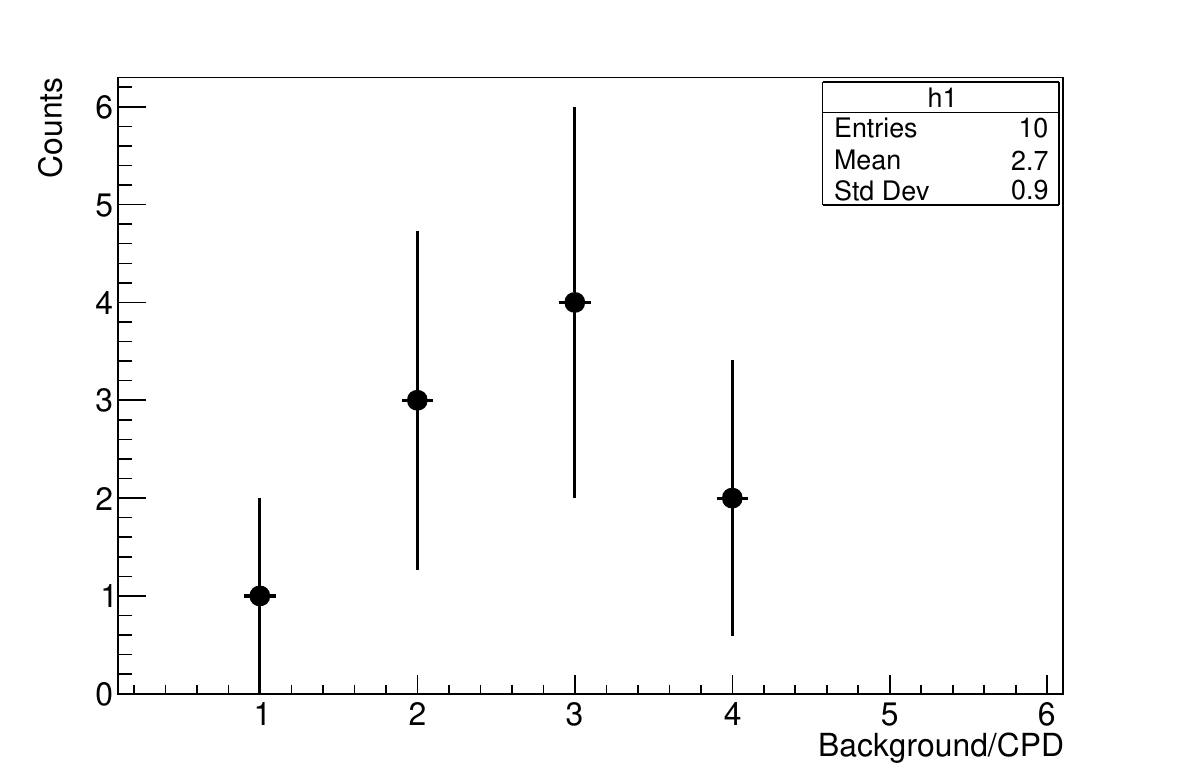}
\caption{Distribution of the background event rate. During the background measurement, the Mn-fiber was sealed for 10 days. The uncertainties on the Y-axis are statistical only.}
\label{fig.Rabackground}
\end{figure}

The system background was determined by measuring 5~g of unused Mn-fiber using the standard sealing and analysis procedure. Ten replicate runs yielded an average background rate of 2.7~CPD (Fig.~\ref{fig.Rabackground}), reflecting contributions from both the apparatus and the Mn-fiber.

\begin{table*}[ht]
\caption{Measurements of $^{226}$Ra concentration (using our custom device), particulate matter count (using a Bettersize C400 optical particle counter), and resistivity (using an AZ8302 portable meter and a +GF+ Signet 9900 in-line meter) were conducted across the 100~t/h UPW system. Uncertainties are reported as follows: for $^{226}$Ra concentration, the statistical uncertainties are based on Poisson statistics, while the systematic uncertainties include contributions from the calibration factor, water volume, and $^{226}$Ra extraction efficiency; for particulate matter count, uncertainties are statistical only; and for resistivity, uncertainties correspond to the instrument display accuracy. Missing resistivity values are due to meter limitations at high resistivity ranges. The large relative uncertainty at S17 for the particulate matter arises from counting statistics: each measurement used a 20~mL sample, yielding an average of 2 counts per 20~mL, which converts to 100 $\pm$ 70~counts/L based on Poisson uncertainty ($\sqrt{2}$ $\times$ 50). For sampling locations, refer to Fig.~\ref{fig.watersystem}.}
\label{tab.Raconcentration}
\centering
\renewcommand{\arraystretch}{1.5}
\begin{tabular}{@{\extracolsep{\fill} } c|c|c|c|c}
\hline
{ \quad No. \quad}&{\qquad $^{226}$Ra concentration\qquad} & {\qquad Particulate matter $>$ 0.5~$\mu$m\qquad} &{ \qquad Resistivity \qquad} &{\qquad Location\qquad} \\
& ($\mu$Bq/m$^3$) & (Counts/L) &  (M$\Omega\cdot$cm) & \\ \hline
1  & (429 $\pm$ 4$_{stat}$ $\pm$ 68$_{sys}$) $\times$ 10$^3$  & 601150 $\pm$ 5482 &  0.019 $\pm$ 0.001 & S1 \\ \hline
2  & (186 $\pm$ 3$_{stat}$ $\pm$ 29$_{sys}$) $\times$ 10$^3$ & 298750 $\pm$ 3864   &0.019 $\pm$ 0.001 & S2 \\\hline
3  & (194 $\pm$ 3$_{stat}$ $\pm$ 31$_{sys}$) $\times$ 10$^3$ &  341850 $\pm$ 4134  & 0.019 $\pm$ 0.001 & S3 \\ \hline
4  & (147 $\pm$ 2$_{stat}$ $\pm$ 23$_{sys}$) $\times$ 10$^3$ & 17650 $\pm$ 939  & 0.019 $\pm$ 0.001 & S4 \\ \hline
5  & (426 $\pm$ 4$_{stat}$ $\pm$ 67$_{sys}$) $\times$ 10$^3$ & 22500 $\pm$ 1060 & 0.018 $\pm$ 0.001 & S5 \\ \hline
6  & 504 $\pm$ 114$_{stat}$ $\pm$ 71$_{sys}$  & 46600 $\pm$ 1526 & 0.018 $\pm$ 0.001 & S6 \\ \hline
7  & (76 $\pm$ 1$_{stat}$ $\pm$ 11$_{sys}$) $\times$ 10$^3$ &  341850 $\pm$ 4134  & 0.018 $\pm$ 0.001 & S7 \\ \hline
8  & (55 $\pm$ 1$_{stat}$ $\pm$ 8$_{sys}$) $\times$ 10$^3$ & 244900 $\pm$ 3499 & 0.052 $\pm$ 0.001 & S8  \\ \hline
9  & 79 $\pm$ 24$_{stat}$ $\pm$ 9$_{sys}$ & 2150 $\pm$ 328 &  0.190 $\pm$ 0.002  & S9 \\ \hline
10 & (15.8 $\pm$ 0.6$_{stat}$ $\pm$ 2.2$_{sys}$) $\times$ 10$^3$ & 10050 $\pm$ 709 &  0.470 $\pm$ 0.005   & S10 \\ \hline
11 & 84 $\pm$ 25$_{stat}$ $\pm$ 17$_{sys}$  & 2750 $\pm$ 370 &  /  & S11 \\ \hline
12 & 96 $\pm$ 26$_{stat}$ $\pm$ 12$_{sys}$ &  2250 $\pm$ 335  &   /  & S12 \\ \hline
13 & 56 $\pm$ 21$_{stat}$ $\pm$ 7$_{sys}$ & 2150 $\pm$ 327 & 17.74 $\pm$ 0.01 & S13 \\ \hline
14 & 49 $\pm$ 22$_{stat}$ $\pm$ 6$_{sys}$ & 900 $\pm$ 200 &  /   & S14 \\ \hline
15 & 6.3 $\pm$ 4.5$_{stat}$ $\pm$ 0.7$_{sys}$ & 850 $\pm$  206 &  18.19 $\pm$ 0.01  & S15 \\ \hline
16 & 19.4 $\pm$ 5.8$_{stat}$ $\pm$ 2.2$_{sys}$  & 350 $\pm$ 132  &  /  & S16 \\ \hline
17 & $<$4 & 100 $\pm$ 70  &   18.20 $\pm$ 0.01   & S17 \\ \hline
\end{tabular}
\end{table*}

Using the parameters described above and applying Eq.~\ref{eq6}, a sensitivity of 3.9~$\mu$Bq/m$^3$ for $^{226}$Ra concentration in water was obtained. This calculation assumes 20~m$^3$ of water passes through 5~g Mn-fiber with a $^{226}$Ra extraction efficiency of 89.2\%, and the data-taking time for $^{222}$Rn activity measurement is 24~h.

\section{Results}\label{sec.Results}
\subsection{$^{226}$Ra concentrations across the system}

Using the custom-developed device described above, the $^{226}$Ra concentrations in water at different locations of the JUNO 100~t/h UPW system, marked as S1--S17 in Fig.~\ref{fig.watersystem}, were measured, and the results are shown in Tab.~\ref{tab.Raconcentration}. The water volume used for S1--S5 was 0.5~m$^3$, for S6--S8 and S10 was 1~m$^3$, for S9 and S11--S14 was 5~m$^3$, and for S15--S17 was 20~m$^3$. All measurements were performed after a 10-day Mn-fiber sealing period and a 24-hour $^{222}$Rn counting interval. Reported uncertainties include both statistical and systematic contributions. During the 45-day filling campaign, a total of six samples were collected at the S17 sampling point, and all measurement results were below the detector's average background; therefore, only an upper limit is provided.

In low-salinity solutions, $^{226}$Ra exists predominantly as uncomplexed Ra$^{2+}$~\cite{bib26,bib27,bib28}, making it removable via deionization processes. According to Tab.~\ref{tab.Raconcentration}, the system reduces $^{226}$Ra concentration from $\sim$4 $\times$ 10$^5$~$\mu$Bq/m$^3$ to $<$4~$\mu$Bq/m$^3$, representing a reduction of over five orders of magnitude. This $>$99.99\% removal efficiency was achieved primarily through the synergistic action of the softening resin and the two-stage RO membranes, with each exhibiting $>$99\% $^{226}$Ra removal efficiency. The $^{226}$Ra removal efficiencies of the main devices used in the UPW system are shown in Tab.~\ref{tab.efficiency}. The slight variation in $^{226}$Ra removal efficiency between the two RO stages is primarily attributable to differences in influent $^{226}$Ra concentration. This system employs two polishing resins in series, with tabulated efficiency calculations assuming identical $^{226}$Ra removal performance.

\begin{table*}[ht]
\caption{$^{226}$Ra removal efficiencies of the devices used in the UPW system. When calculating the $^{226}$Ra removal efficiency of the polishing resin, it is assumed that both resin stages have the same efficiency.}
\label{tab.efficiency}
\renewcommand{\arraystretch}{1.3}
\begin{tabular} {@{\extracolsep{\fill} } c|c|c|c}
\toprule
{No.}&{ Device} &{$^{226}$Ra removal efficiency (\%) } & {$^{226}$Ra concentration before the device ($\mu$Bq/m$^3$)}\\ \hline
1   & Softening resin  & 99.88 $\pm$ 0.03$_{stat}$ $\pm$ 0.02$_{sys}$ & $\sim$4.3 $\times$ 10$^5$\\ \hline
2   & 1$^{st}$ stage of RO & 99.86 $\pm$ 0.04$_{stat}$ $\pm$ 0.03$_{sys}$ & $\sim$5.5 $\times$ 10$^4$\\ \hline
3   & 2$^{nd}$ stage of RO & 99.47 $\pm$ 0.16$_{stat}$ $\pm$ 0.13$_{sys}$ & $\sim$1.6 $\times$ 10$^4$\\ \hline
4   &  Polishing resin & 64.1 $\pm$ 15.1$_{stat}$ $\pm$ 3.0$_{sys}$ & $\sim$49\\ \hline
5   &  EDI &  Below reliable quantification limit  & / \\ \hline
\end{tabular}
\end{table*}

However, according to the data shown in Tab.~\ref{tab.Raconcentration}, mechanical filtration units (e.g., bag filters, cartridges, multi-media filters) also exhibited incidental $^{226}$Ra removal via adsorption of Ra$^{2+}$ onto particulates, enabling concurrent removal during filtration. Among all devices, only RO integrates both deionization and particulate removal~\cite{bib29}.

In addition, the data in Tab.~\ref{tab.Raconcentration} also indicates that the $^{226}$Ra concentration in water increased after passing through each tank, which is primarily attributable to air ingress through tank ventilation valves. Fiberglass-reinforced plastic tanks require pressure-equalizing vent valves to prevent structural damage during water level fluctuations. These valves introduce ambient air (unfiltered for aboveground tanks, filtered for underground tanks) containing dust-bound $^{226}$Ra progeny that dissolve upon water contact. A significant increase in $^{226}$Ra concentration after the pretreatment water tank was traced to sediment introduced during initial commissioning, when sea salt was used for softening resin regeneration. Although sea salt was replaced by clean regeneration salt and the tank was flushed, residual sediment on the tank walls persisted as a leaching source.

\subsection{Relationship between $^{226}$Ra concentration and particulate counts and resistivity}
Tab.~\ref{tab.Raconcentration} also lists the particulate counts and resistivity values across the water system. The particulate counts were measured using a Bettersize C400 optical particle counter (Dandong Baxter); the small sample volume (20~mL) resulted in significant data uncertainties. Resistivity was measured using an AZ8302 portable conductivity tester (Taiwan Hengxin) upstream of the secondary RO and a +GF+ Signet 9900 in-line meter downstream. Portable measurements become unreliable when the resistivity exceeds 10~M$\Omega\cdot$cm, and the in-line measurements prevent modification of the sampling point. These constraints resulted in missing high-resistivity data at multiple locations.

The observed correlations among $^{226}$Ra concentration, particulate counts, and resistivity are a concurrent outcome of the multi-stage purification process. The softening resin achieves $>$99\% $^{226}$Ra removal by ion exchange (Na$^{+}$ replaces Ra$^{2+}$), which minimally influences resistivity and does not filter particles due to its 100--500~$\mu$m bead size. The simultaneous decrease in all parameters is attributable to their collective removal through successive treatment stages.

\section{Discussion} \label{sec.discussion}
The system reduces $^{226}$Ra concentration in water from $\sim$4 $\times$ 10$^5$~$\mu$Bq/m$^3$ to $<$4~$\mu$Bq/m$^3$, corresponding to a total reduction exceeding five orders of magnitude and an overall removal efficiency of $>$99.999\%. Even though the RO provides high removal efficiency for particulate matter and ions, a quantitative breakdown of the contributions from individual components (Tab.~\ref{tab.efficiency}) reveals that the softening resin plays a critical and complementary role, accounting for a 99.88 $\pm$ 0.04\% reduction alone. Our resin screening revealed significant variation in $^{226}$Ra removal efficiency, with cationic resins demonstrating superior performance over mixed-bed (anionic/cationic) alternatives. The observed difference is attributed to the dominant speciation of $^{226}$Ra as Ra$^{2+}$ in low-salinity water~\cite{bib26,bib27,bib28}, which favors exchange with cationic functional groups. This mechanistic understanding guided the selection of a dedicated cationic softening resin over conventional mixed-bed polishing resins. Although the selected softening resin achieves high $^{226}$Ra rejection ($>$99\%), its pre-RO positioning is mandatory and not interchangeable with post-RO polishing resin due to substantial organic leaching. These leachates cause measurable deterioration of water quality, primarily through resistivity reduction. Quantitative water quality analysis showed that placing the polishing resin downstream of the RO caused resistivity to drop from $\sim$18~M$\Omega\cdot$cm to below 1~M$\Omega\cdot$cm, confirming that organic leachates from the resin degrade water quality if placed after the RO and justifying the current upstream configuration.

In reducing the source of $^{226}$Ra pollution, the following two contamination sources were identified:

(1) Thorough cleaning of the pretreatment water tank. The periodic regeneration of softening resin, commonly performed with economical sea salt, was found to introduce sediment containing significant $^{226}$Ra contamination. During the initial commissioning phase of the water system, conventional sea salt was used, resulting in a sharp increase in $^{226}$Ra concentration --- from $\sim$500~$\mu$Bq/m$^3$ to $\sim$3.0 $\times$ 10$^5$ $\mu$Bq/m$^3$ --- after water passed through the pretreatment tank. Visual inspection revealed a substantial layer of sediment at the tank bottom, which subsequent analysis confirmed originated from the sea salt. The data presented in this paper were collected after switching to clean regenerant salt and performing systematic flushing of the tank. The reduction from $\sim$3.0 $\times$ 10$^5$~$\mu$Bq/m$^3$ to $\sim$7.6 $\times$ 10$^4$~$\mu$Bq/m$^3$ confirms the effectiveness of replacing the regenerant salt and cleaning the tanks, representing an additional 74.7\% reduction in post-tank concentration.

However, even after adopting clean salt and flushing, residual sediment remains adhered to the tank walls. This residual contamination accounts for the persistent $^{226}$Ra increase of $\sim$7.6 $\times$ 10$^4$~$\mu$Bq/m$^3$ observed after the pretreatment water tank. To fully eliminate this source of contamination, a thorough cleaning of the tank interior is required. As this procedure entails draining the water tank and a temporary system shutdown, it will be scheduled for a suitable maintenance window in the future.

(2) Mitigation of contamination from storage tank breather valves. Storage tanks introduce measurable radiological contamination primarily through their air-admittance breather valves. Comparative measurements at sampling points S9 (upstream of the tank) and S10 (downstream of the tank) show that the $^{226}$Ra concentration increased from (79 $\pm$ 24$_{stat}$ $\pm$ 9$_{sys}$)~$\mu$Bq/m$^3$ to (15.8 $\pm$ 0.6$_{stat}$ $\pm$ 2.2$_{sys}$) $\times$ 10$^3$~$\mu$Bq/m$^3$ after passing through the tank, an increase of 200 times. This sharp rise confirms that air ingress via the breather valve is the dominant contamination mechanism. Two mitigation strategies have been evaluated: installing particulate-retentive filters on breather assemblies or using boil-off nitrogen as a clean breathing gas. While nitrogen blanketing could thoroughly eliminate background contamination, it would require a dedicated gas supply system at high cost. Furthermore, the existing boil-off nitrogen capacity at the JUNO site is insufficient to meet the system's operational demand. Therefore, installing filtration systems at each tank air inlet has been selected as the practical solution, which will be implemented during a suitable maintenance period.

\section{Summary}\label{sec.summary}
JUNO is a multi-purpose neutrino experiment primarily designed to determine the neutrino mass ordering. To ensure detector safety and minimize background, the CD must be filled with UPW containing $<$50~$\mu$Bq/m$^3$ of $^{226}$Ra before LS introduction -- a requirement that presents substantial technical challenges given the ultra-low concentration target and the large water volume involved.

A dedicated 100~t/h UPW production and circulation system was developed, integrating multi-stage purification --- including specialized softening resin and RO --- with a novel, high-sensitivity $^{226}$Ra measurement system (3.9~$\mu$Bq/m$^3$) based on Mn-fiber adsorption and radon emanation. Systematic characterization across the purification stages revealed $>$99\% removal efficiency for both the softening resin and the two-stage RO. Quantitative analysis further identified regenerant salt impurities and air ingress via tank breather valves as the two dominant contamination pathways, accounting for the majority of post-purification recontamination.

The JUNO 100~t/h water system successfully reduced $^{226}$Ra concentration from $\sim$4 $\times$ 10$^5$~$\mu$Bq/m$^3$ to $<$4~$\mu$Bq/m$^3$, achieving a reduction exceeding five orders of magnitude and surpassing the JUNO requirement by more than an order of magnitude. This performance establishes a world-leading benchmark in ultra-low-$^{226}$Ra water purification.

Looking forward, the identified contamination pathways provide a clear basis for further optimization. Planned operations --- including thorough cleaning of the pretreatment tank and installation of high-efficiency particulate filters on the tank breather valve --- are expected to further reduce the $^{226}$Ra concentration in water. More broadly, the developed high-sensitivity measurement and multi-stage purification strategy can serve as a blueprint for future low-background experiments, such as Hyper-K~\cite{bib30}, which requires stringent radiopurity control of large-scale water systems.


\begin{thebibliography}{99}

\bibitem{bib1}  F. An, G. An, Q. An et al., Neutrino physics with JUNO. J. Phys. G {\bf 43}, 030401 (2016). \href{https://iopscience.iop.org/article/10.1088/0954-3899/43/3/030401}{https://doi.org/10.1088/0954-3899/43/3/030401}
\bibitem{bib2}  A. Abusleme, T. Adam, S. Ahmad et al., JUNO physics and detector. Prog. Part. Nucl. Phys. {\bf 123}, 103927 (2022). \href{https://doi.org/10.1016/j.ppnp.2021.103927}{https://doi.org/10.1016/j.ppnp.2021.103927}
\bibitem{bib2-0-1} S. Zhang, Y.B. Huang, M. He et al., Sub-GeV events energy reconstruction with 3-inch PMTs in JUNO, Nucl. Sci. Tech. 36, 84 (2025). \href{https://doi.org/10.1007/s41365-025-01678-4}{https://doi.org/10.1007/s41365-025-01678-4}
\bibitem{bib2-0-2} M.H. Liao, K.X. Huang, Y.M. Zhang et al., A ROOT-based detector geometry and event visualization system for JUNO-TAO. Nucl. Sci. Tech. 36, 39 (2025). \href{https://doi.org/10.1007/s41365-024-01604-0}{https://doi.org/10.1007/s41365-024-01604-0}

\bibitem{bib3} Junwei Zhang, Yongpeng Zhang, Yongbo Huang et al., Reconstruction of the effective energy-deposition vertex of muon showers using PMT waveform in a large-scale liquid scintillator detector, Eur. Phys. J. C 86:306 (2026). \href{https://doi.org/10.1140/epjc/s10052-026-15536-z}{https://doi.org/10.1140/epjc/s10052-026-15536-z}

\bibitem{bib4}  A. Abusleme, T. Adam, S. Ahmad et al., Radioactivity control strategy for the JUNO detector. J. High Energy Phys. {\bf 2021}, 102 (2021). \href{https://doi.org/10.1007/JHEP11(2021)102}{https://doi.org/10.1007/JHEP11(2021)102}

\bibitem{bib2-2} J. Boger, R.L Hahn, J.K. Rowley et al., The Sudbury Neutrino Observatory, Nucl. Instrum. Meth. A {\bf 449}, 172-207 (2000). \href{https://doi.org/10.1016/S0168-9002(99)01469-2}{https://doi.org/10.1016/S0168-9002(99)01469-2}

\bibitem{bib2-3} S. Fukuda, Y. Fukuda, T. Hayakaw et al., The Super-Kamiokande Detector, Nucl. Instrum. Meth. A {\bf 501}, 418-462 (2003). \href{https://doi.org/10.1016/S0168-9002(03)00425-X}{https://doi.org/10.1016/S0168-9002(03)00425-X}

\bibitem{bib14}  T.C. Andersen, I. Blevis, J. Boger et al., Measurement of radium concentration in water with Mn-coated beads at the Sudbury Neutrino Observatory. Nucl. Instrum. Meth. A {\bf 501}, 399-417 (2003). \href{ http://dx.doi.org/10.1016/S0168-9002(03)00616-8}{https://doi.org/10.1016/S0168-9002(03)00616-8}

\bibitem{bib2-4} T. Ueno, Radium Concentration Measurements in Super-Kamiokande and Search for Magnetic Monopoles in Grand Unified Theory [Master's Thesis], University of Tokyo (2021). \href{https://www-sk.icrr.u-tokyo.ac.jp/assets/paper/list/pdf/mthesis/2008/master_ueno.pdf}{https://www-sk.icrr.u-tokyo.ac.jp/assets/paper/list/pdf/mthesis/2008/master-ueno.pdf}

\bibitem{bib5} Standardization Administration of China. GB/T 11446.1-2013, Electronic grade water. Beijing: Standards Press of China, 2013. \href {https://std.samr.gov.cn/}{https://std.samr.gov.cn/}

\bibitem{bib6}  M. Tawalbeh, L. Qalyoubi, A.Al-Othman et al., Insights on the development of enhanced antifouling reverse osmosis membranes: Industrial applications and challenges. Desalination {\bf 553}, 116460 (2023). \href{https://doi.org/10.1016/j.desal.2023.116460}{https://doi.org/ 10.1016/j.desal.2023.116460}

\bibitem{bib7} Z.U. Khan, M. Moronshing, M. Shestakova et al., Electro-deionization (EDI) technology for enhanced water treatment and desalination: A review. Desalination {\bf 548}, 116254 (2023). \href{https://doi.org/10.1016/j.desal.2022.116254}{https://doi.org/10.1016/j.desal.2022.116254}

\bibitem{bib8}  B. Wang, C. Guo, Y.G. Xie et al., Measurement of the radon concentration in the ultrapure water of the JUNO-LS UPW system. J. Instrum. {\bf 20}, P03028 (2025). \href{https://doi.org/10.1088/1748-0221/20/03/P03028}{htps://doi.org/10.1088/1748-0221/20/03/P03028}

\bibitem{bib9}  Y. Nakano, T. Hokama, M. Matsubaara et al., Measurement of the radon concentration in purified water in the Super-Kamiokande IV detector. Nucl. Instrum. Meth. A {\bf 977}, 164297 (2020). \href{https://doi.org/10.1016/j.nima.2020.164297}{https://doi.org/ 10.1016/j.nima.2020.164297}

\bibitem{bib10}  T.Y. Guan, Y. P. Zhang, B. Wang et al., Development of low-radon ultra-pure water for the Jiangmen Underground Neutrino Observatory. Nucl. Instrum. Meth. A {\bf 1063}, 169244 (2024). \href{https://doi.org/10.1016/j.nima.2024.169244}{https://doi.org/ 10.1016/j.nima.2024.169244}

\bibitem{bib11}  G.R. Araujo, L. Baudis, Y. Biondi et al., The upgraded low-background germanium counting facility Gator for high-sensitivity $\gamma$-ray spectrometry. J. Instrum. {\bf 17}, P08010 (2022). \href{https://doi.org/10.1088/1748-0221/17/08/P08010}{https://doi.org/10.1088/1748-0221/17/08/P08010}

\bibitem{bib12}  S. Ito, Y. Takaku, M. Ikeda et al., Determination of trace level thorium and uranium in high purity gadolinium sulfate using ICP-MS with solid-phase chromatographic extraction resins. AIP Cong. Proc. {\bf 1921}, 030003 (2018). \href{https://doi.org/10.1063/1.5018990}{https://doi.org/ 10.1063/1.5018990}

\bibitem{bib13} W.S. Moore and D.F. Reid, Extraction of radium from natural waters using manganese-impregnated acrylic fibers. J. Geophys. Res. {\bf 78}, 8880-8886 (1973). \href{http://dx.doi.org/10.1029/JC078i036p08880}{https://doi.org/10.1029/JC078i036p08880}

\bibitem{bib15}  R.N. Peterson, W.C. Burnett, N. Dimova et al., Comparison of measurement methods for radium-226 on manganese-fiber. Limnol. Oceanogr. Methods{\bf 7}, 196-205 (2009). \href{https://doi.org/10.4319/lom.2009.7.196}{https://doi.org/10.4319/lom.2009.7.196}

\bibitem{bib16}  L.F. Xie, J.C. Liu, S.K. Qiu et al., Developing the radium measurement system for the water Cherenkov detector of the Jiangmen Underground Neutrino Observatory. Nucl. Instrum. Meth. A {\bf 976}, 164266 (2020). \href{http://dx.doi.org/10.1016/j.nima.2020.164266}{https://doi.org/10.1016/j.nima.2020.164266}

\bibitem{bib17}  C. Li, B. Wang, Y. Liu et al., Developing a $\mu$Bq/m$^3$ level $^{226}$Ra concentration in water measurement system for the Jiangmen Underground Neutrino Observatory. Nucl. Instrum. Meth. A {\bf 1063}, 169257 (2024). \href{https://doi.org/10.1016/j.nima.2024.169257}{https://doi.org/ 10.1016/j.nima.2024.169257}

\bibitem{bib19}  J.X. Wang, T.C. Andersen, J.J. Simpson, An electrostatic radon detector designed for water radioactivity measurements. Nucl. Instrum. Meth. A {\bf 421}, 601-609 (1999). \href{http://dx.doi.org/10.1016/S0168-9002(98)01230-3}{https://doi.org/10.1016/S0168-9002(98)01230-3}

\bibitem{bib20}  Y.Y. Chen, Y.P. Zhang, Y. Liu et al., A study on the radon removal performance of low background activated carbon. J. Instrum. {\bf 17}, P02003 (2022). \href{http://dx.doi.org/10.1088/1748-0221/17/02/P02003}{https://doi.org/10.1088/1748-0221/17/02/P02003}

\bibitem{bib21}  C. Guo, J.C. Liu, Y.P. Zhang et al., Study on the radon removal for the water system of Jiangmen Underground Neutrino Observatory. Radiat. Detect. Technol. Methods {\bf 2}, 48 (2018). \href{http://dx.doi.org/10.1007/s41605-018-0077-8}{https://doi.org/10.1007/s41605-018-0077-8}

\bibitem{bib22}  Y. Liu, Y.P. Zhang, J.C. Liu et al., System upgrade for $\mu$Bq/m$^3$ level $^{222}$Rn concentration measurement. J. Instrum. {\bf 18}, T03002 (2023). \href{http://dx.doi.org/10.1088/1748-0221/18/03/T03002}{https://doi.org/10.1088/1748-0221/18/03/T03002}

\bibitem{bib23}  C. Li, Y. Zhang, L. Lv et al., Study on the radon adsorption capability of low-background activated carbon. J. Radioanal. Nucl. Chem. {\bf 333}, 337-346 (2024). \href{http://dx.doi.org/10.1007/s10967-023-09211-w}{https://doi.org/10.1007/s10967-023-09211-w}

\bibitem{bib24}  Y. Nakano, H. Sekiya, S. Tasaka et al., Measurement of radon concentration in super-Kamiokande's buffer gas. Nucl. Instrum. Meth. A {\bf 867}, 108-114 (2017). \href{ http://dx.doi.org/10.1016/j.nima.2017.04.037}{https://doi.org/10.1016/j.nima.2017.04.037}

\bibitem{bib25} C.B.Z. Luo, Q. Tang, C. Guo et al., Measurement of radon concentration in the output water of the 100 t/h ultrapure water system at the Jiangmen Underground Neutrino Observatory, Nucl. Instrum. Meth. A {\bf 1087}, 171428 (2026). \href{https://doi.org/10.1016/j.nima.2026.171428}{https://doi.org/10.1016/j.nima.2026.171428}

\bibitem{bib25-2} P.N. Bhattiprolu1, S.P. Martin, J.D. Wells, Criteria for projected discovery and exclusion sensitivities of counting experiments, Eur. Phys. J. C 81:123 (2021). \href{https://doi.org/10.1140/epjc/s10052-020-08819-6}{https://doi.org/10.1140/epjc/s10052-020-08819-6}

\bibitem{bib26}B.L. Dickson, The environmental behavior of radium. IAEA. Vol. 1. No. 310. 335-372 (1990). \href{https://inis.iaea.org/records/swprd-wbq70/files/21039568.pdf}{https://inis.iaea.org/records/swprd-wbq70/files/21039568.pdf}

\bibitem{bib27}  D. Porcelli and P.W. Swarzenski, The Behavior of U- and Th-series Nuclides in Groundwater. Rev. Mineral. Geochem. {\bf 52}, 317-361 (2003). \href{https://doi.org/10.2113/0520317}{https://doi.org/ 10.2113/0520317}

\bibitem{bib28}  F. Carvalho, D. Chambers, S. Fernandes et al., The Environmental Behaviour of Radium: Revised Edition. IAEA Technical Reports Series No. 476 (2014). \href{https://www.iaea.org/publications/10478}{https://www.iaea.org/publications/10478}

\bibitem{bib29} PURETEC Industrial Waterm, The Basics - Reverse Osmosis. \href{https://puretecwater.com/wp-content/uploads/2024/04/Reverse-Osmosis-The-Basics-032024-1.pdf}{https://puretecwater.com}

\bibitem{bib30} K. Abe and H.K. Tanaka, Hyper-Kamiokande construction status and prospects. Front. Phys. 12:1378254 (2024). \href{https://doi.org/10.3389/fphy.2024.1378254}{ https://doi.org/10.3389/fphy.2024.1378254}

\end{thebibliography}
\end{document}